\documentclass[onecolumn, 12pt, journal]{IEEEtran}

\usepackage[utf8]{inputenc}
\usepackage{amsmath}
\usepackage{amssymb}
\newcommand{\R}{\mathbb{R}}
\newcommand{\C}{\mathbb{C}}
\newcommand{\ra}{\rightarrow}

\usepackage{multirow}
\usepackage[pdftex]{graphicx}
\usepackage{multirow,setspace}
\usepackage{subcaption}
\usepackage{balance}
\usepackage{bm}

\usepackage{epsfig,color}
\usepackage{epstopdf}

\usepackage{cite}

\begin{document}

\bstctlcite{IEEEexample:BSTcontrol}

\title{Graph Signal Processing: Vertex Multiplication}

\author{Aykut~Ko\c{c} and~Yigit E. Bayiz
\thanks{Yigit E. Bayiz is with the Department of Electrical and Electronics Engineering, Bilkent University, Bilkent, Ankara, Turkey}
\thanks{Aykut Ko\c{c} (Senior Author) is with the Department of Electrical and Electronics Engineering, Bilkent University, Bilkent, Ankara, Turkey, and also with the National Magnetic Resonance Research Center (UMRAM), Bilkent University, Bilkent, Ankara, Turkey, e-mail: aykut.koc@bilkent.edu.tr}}

\markboth{}%
{Shell \MakeLowercase{\textit{et al.}}: Bare Demo of IEEEtran.cls for IEEE Journals}

\maketitle

\begin{abstract}
On the Euclidean domains of classical signal processing, linking of signal samples to the underlying coordinate structure is straightforward. While graph adjacency matrices totally define the quantitative associations among the underlying graph vertices, a major problem in graph signal processing is the lack of explicit association of vertices with an underlying quantitative coordinate structure. To make this link, we propose an operation, called the \emph{vertex multiplication}, which is defined for graphs and can operate on graph signals. Vertex multiplication, which generalizes the coordinate multiplication operation in time series signals, can be interpreted as an operator which assigns a coordinate structure to a graph. By using the graph domain extension of differentiation and graph Fourier transform (GFT), vertex multiplication is defined such that it shows Fourier duality, which states that differentiation and coordinate multiplication operations are duals of each other under Fourier transformation (FT). The proposed definition is shown to reduce to coordinate multiplication for graphs corresponding to time series. Numerical examples are also presented.
  
\end{abstract}


 \ifCLASSOPTIONpeerreview
 \begin{center} \bfseries EDICS Category: 3-BBND \end{center}
 \fi
%
\IEEEpeerreviewmaketitle

\section{Introduction}

Classical digital signal processing (DSP) provides a useful tool for the analysis of signals defined by sampling an Euclidian space such as time-series signals and the $\R^2$ plane for images. However, the classical DSP theory is not designed to capture the complicated structures of large networks, such as social and economic networks, networks arising from the world wide web, and sensor networks. The graph signal processing (GSP) provides a new framework that can make the analysis of these large networks possible,~\cite{shuman13emerging,ortega18gspsurvey,
ribeiro18gsp,stankovic19gsp1,stankovic19gsp2,gavili17shiftoperator,sandryhaila14bigdata,sandryhaila14freq,stankovic19vertexfreq,chen15sampling,wang17gspfrt,wang18gspfrtsampling,sandryhaila13filtersicassp,sandryhaila13discretegsp}.  
The GSP framework aims towards developing efficient methods for analyzing and processing signals with complex underlying structures such as networks or graphs \cite{ortega18gspsurvey,sandryhaila13discretegsp}. Graph based extensions to several other areas such as machine learning and analysis of brain signals have also been investigated \cite{stankovic20gsp3machinelearning,huang16graphbrain}.  

Classical signal processing concepts have been extended to the graph domain \cite{ortega18gspsurvey,ribeiro18gsp,stankovic19gsp1}. Specifically, filtering \cite{stankovic19gsp2,ribeiro18gsp,shuman13emerging,
sandryhaila13filtersicassp,ortega18gspsurvey,
gavili17shiftoperator,sandryhaila13discretegsp,hua19graphfilter}, frequency analysis \cite{sandryhaila14freq,stankovic19gsp2,grassi17timevertex}, sampling \cite{chen15sampling,anis14sampling}, interpolation \cite{narang13interpolation}, Fourier transform (FT) \cite{sandryhaila13gfticassp}, signal reconstruction \cite{brugnoli20recons}, processing of stationary signals and processes \cite{perraudin17stationary,marques17stationary}, and multiscale decomposition methods \cite{zheng16multiscaledec} have been considered for graph signals.

Much of the developments in the processing of graph signals rely on extending the definition of FT and frequency analysis to graph signals\cite{sandryhaila14freq}. This allows the translation of many signal processing algorithms to graph signals, and provided a framework that has proved to be foundational in many novel GSP applications, including filtering, sampling and interpolation theory,\cite{chen15sampling,narang13interpolation,anis14sampling}, big data analysis, \cite{sandryhaila14bigdata}, and classification, \cite{chen14semisuper,chen13adaptive}. There are two main approaches for extending the FT to the graph domain. The first is derived from the \emph{spectral graph theory} and uses the graph Laplacian. This framework describes the graph Fourier transform (GFT) as a change of basis into the basis of the eigenvectors of the graph Laplacian \cite{shuman13emerging}.
Although this approach is generally considered to be limited to analysis of undirected graphs, there also exist extensions to directed graphs,~\cite{singh16directedlaplacian}.
Built upon the \textit{algebraic signal processing}, the second approach is based on the adjacency matrix of the graph and describes the GFT as a change of basis into the eigenvectors of the adjacency matrix \cite{sandryhaila13discretegsp}. This reflects the intuition that the adjacency matrix is analogous to the discrete shift matrix, and the eigenvalues of the latter form a basis for the discrete Fourier transform (DFT). The second approach, which we have adopted throughout this letter, supports both directed and undirected graphs.

The relationships between operators on graphs and their FT counterparts have also been studied, \cite{sandryhaila13discretegsp,shuman13emerging,ortega18gspsurvey,sandryhaila14freq,stankovic19gsp2,grassi17timevertex
}. Specifically, important operations like convolution, translation, modulation, dilation, and filtering have been generalized to GSP domain in \cite{shuman13emerging}. In \cite{agaskar12uncertaintyongraphs}, the uncertainty relation has been generalized to signals defined on graphs. FT relations for some of the familiar operators such as shift, modulation and dilation have also been studied in \cite{shuman13emerging}. Yet, Fourier duality for the differentiation on graphs has not been addressed.

In this letter, by using the Fourier duality, we propose an operation, called the \emph{vertex multiplication (VM)} for graphs. VM mimics the coordinate multiplication operator of time series signals (${\cal U} f(u) = uf(u)$), which is the Fourier dual of differentiation. While discretization of coordinate multiplication is straightforward, generalization to the graph domain is problematic since the vertices of a graph do not correspond to certain quantitative values, apart from just indices of order. Defined in a matrix form, VM can be interpreted as an operator which assigns a coordinate structure to a graph by associating a \textit{``coordinate vector"} (represented by the columns of the matrix) for each vertex, and can also operate on graph signals (represented as vectors) through matrix multiplication. Since several established metrics and quantitative manipulations can be applied to coordinate vectors, the proposed direct assignment of coordinate structure to the vertex domain can also be instrumental in efforts to define ``distance" metrics in the vertex domain,~\cite{ron11relaxationdistance,lafon06diffusiondistance}, study the notion of smoothness of signals on graphs,~\cite{zhu12apprxgraphsignals}, localization of signals in the vertex-domain and study of transforms on graphs, \cite{coifman06diffusionwavelet,hammond11graphwavelets,narang12waveletgraph,shuman13emerging}.

The proposed VM generalizes a fundamental operation like coordinate multiplication to graph domain and defines the Fourier duality of differentiation for GSP. Moreover, VM, being defined totally consistent with the circulant and dual structure of the DFT, can also be considered a natural way to overcome a major obstacle in embedding the underlying structure of irregular vertex domain to a quantitative coordinate structure assigned to the vertices. This coordinate association is important in the ongoing generalizations from DSP to GSP.

\section{Duality Relation}
The duality between differentiation and coordinate multiplication operators is particularly important in classical signal processing \cite{papoulis77book,cohentannoudji77quantum}. This duality between time (space) and frequency (spatial-frequency) domains is also one of the most fundamental properties of the FT. In mathematical terms, let $\mathcal{U}$, $\mathcal{D}$ and $\mathcal{F}$ denote the coordinate multiplication, differentiation and FT operators, respectively. Continuous manifestations of the former two are:
\begin{equation}
\label{eq:operatorU}
{\cal U} f(u) = uf(u), \\
\end{equation}
\begin{equation}
\label{eq:operatorD}
{\cal D} f(u) = \frac{1}{2\pi j} \frac{df(u)}{du},
\end{equation}
where $(2\pi j)^{-1}$ is included to make ${\cal U}$ and ${\cal D}$ precise Fourier duals (the effect of either in one domain is its dual in the other domain). Then, the duality is given as: 
\begin{equation} \label{U_fdf}
\mathcal{U}=\mathcal{FDF}^{-1}.
\end{equation}
This relation can also manifest itself between shift and modulation operations, both of which are fundamental properties of Fourier analysis,\cite{shuman13emerging}. The duality relations of the FT in the classical signal processing theory are crucial for understanding much of the underlying theory as well as for being instrumental in applications, \cite{koc19ieee}. Therefore, an extension of these relations to GSP is inescapable. The duality relation creates a way to define the coordinate multiplication operator without needing coordinates to be explicitly defined once a differential operator on graphs and GFT are provided. Therefore, the extension of Fourier duality to graphs can be used to define a coordinate structure on graphs where one replaces $\mathcal{D}$ with the differential operator on graphs and $\mathcal{F}$ with GFT.

\section{GFT and Differentiation on Graphs}
A finite \textit{graph} $G = (\mathcal{V},\mathbf{A})$ is a finite set of $N$ ordered points $\mathcal{V} = \{v_0,v_1\dots v_{N-1}\}$ (called vertices) which are connected to each other according to some relation. The connections in $G$ are represented by an \emph{adjacency matrix} $\mathbf{A}$. The element $a_{ij}$ of $\mathbf{A}$ is the weight of the connection between $i$'th vertex and $j$'th vertex where $i,j = 0,1,...,N-1$. In general, this connection is directed. However for undirected graphs for which the connections are not directed, we have a symmetric $\mathbf{A}$ with $ a_{ij} = a_{ji}$.  

Any complex valued function $x$ defined on the set of vertices $\mathcal{V}$, i.e.: $ x:\mathcal{V} \ra \C $
is called a \emph{graph signal}. Since $\mathcal{V}$ is finite, it is convenient to represent $x$ as a vector where each index of $x$ is the value the signal takes on the corresponding vertex:
\begin{equation}
    \mathbf{x} = [x_0,x_1,\dots,x_{N-1}]^{\top},\quad x_i = x(v_i).
\end{equation}

\begin{figure*}[ht]
\begin{subfigure}{.23\textwidth}
  \centering
  \includegraphics[height=0.38\textheight,width=\linewidth,trim={0 3em 0 0},clip]{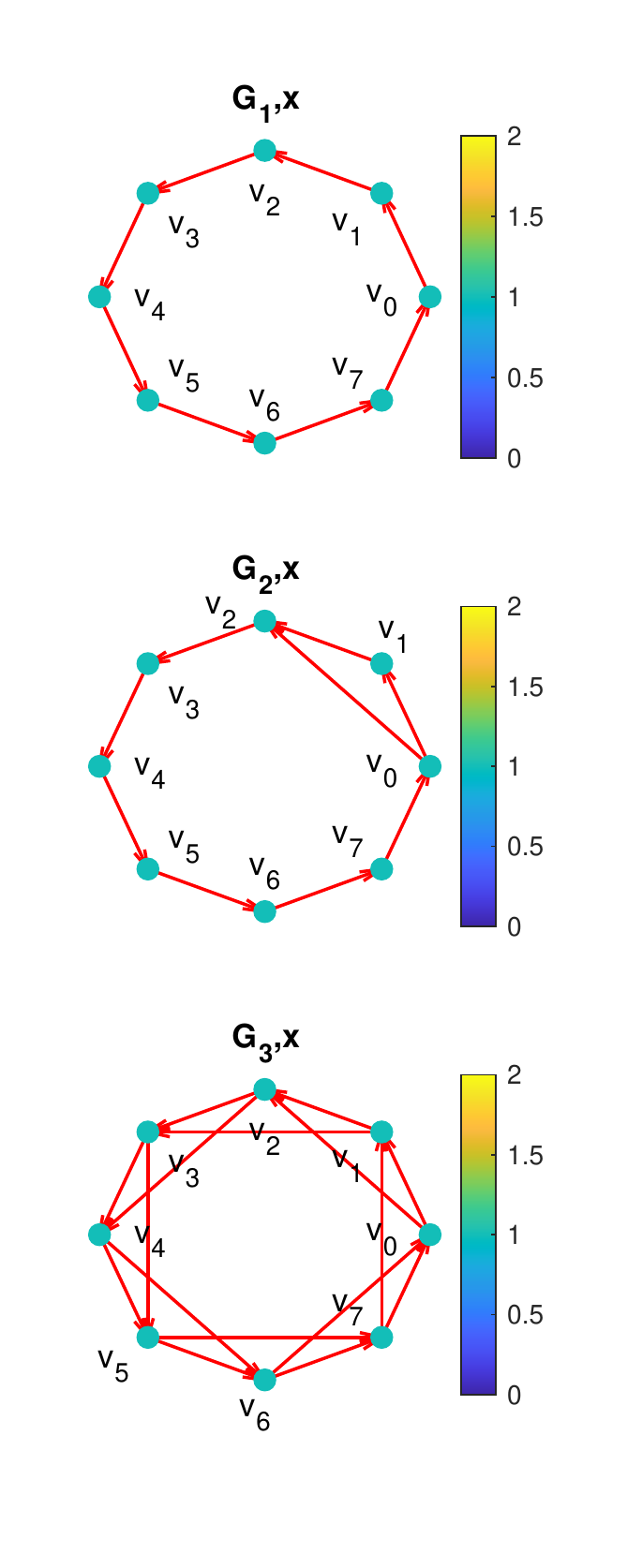} 
  \caption{}
  \label{fig:exampleA}
\end{subfigure}
\begin{subfigure}{.23\textwidth}
  \centering
  \includegraphics[height=0.38\textheight,width=\linewidth,trim={0 3em 0 0},clip]{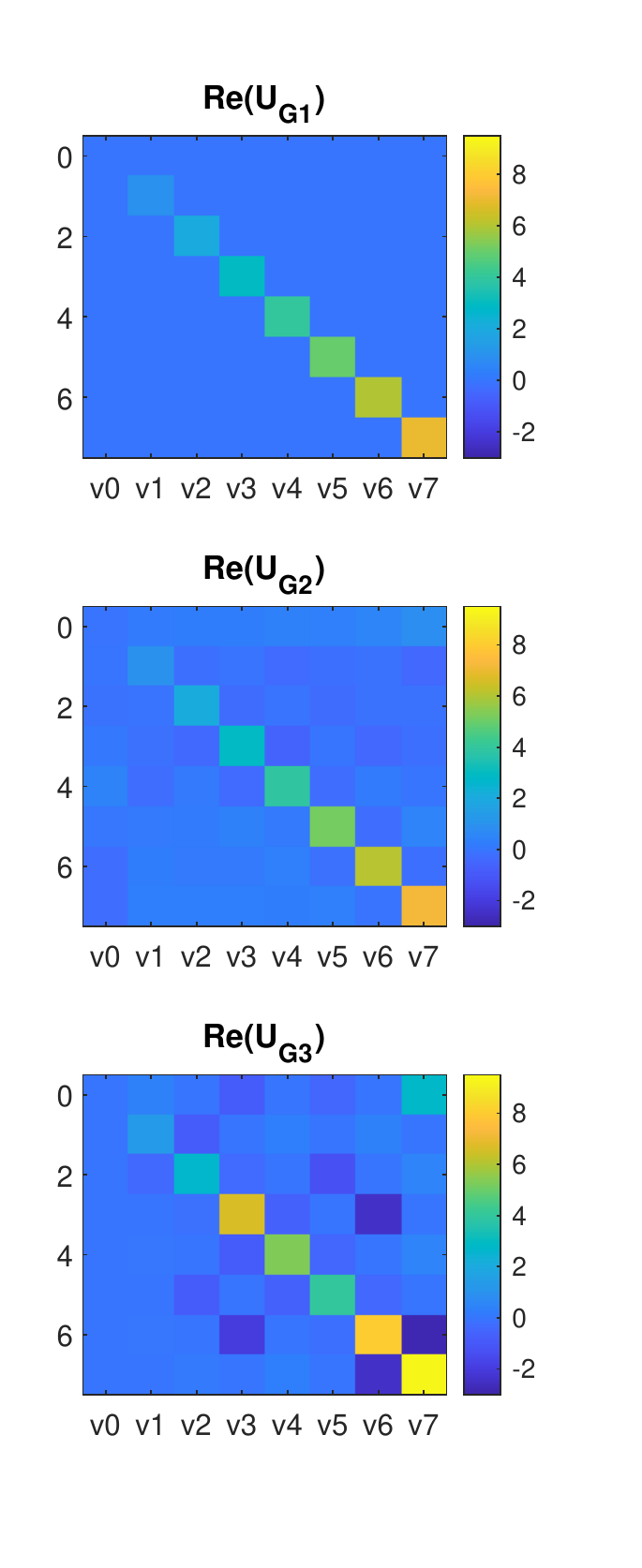} 
  \caption{}
  \label{fig:exampleB}
\end{subfigure}
\begin{subfigure}{.23\textwidth}
  \centering
  \includegraphics[height=0.38\textheight,width=\linewidth,trim={0 3em 0 0},clip]{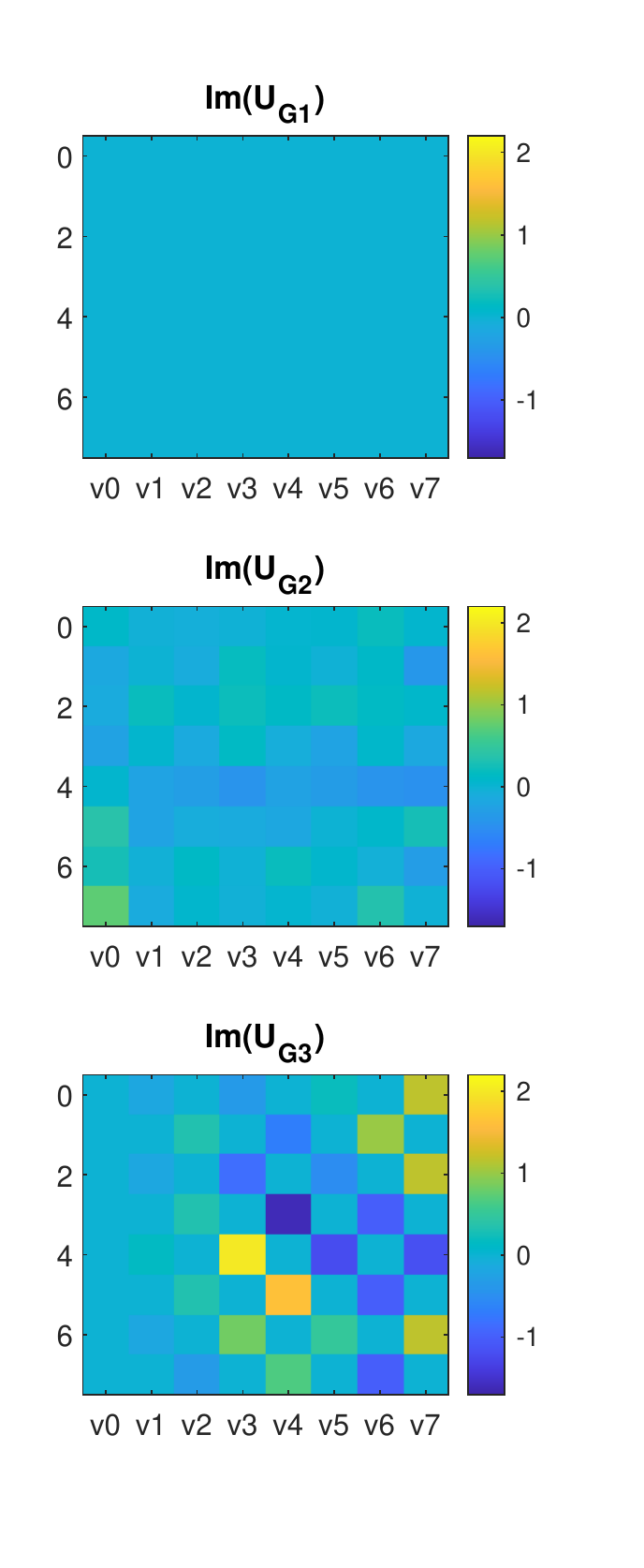} 
  \caption{}
  \label{fig:exampleC}
\end{subfigure}
\begin{subfigure}{.23\textwidth}
  \centering
  \includegraphics[height=0.38\textheight,width=\linewidth,trim={0 3em 0 0},clip]{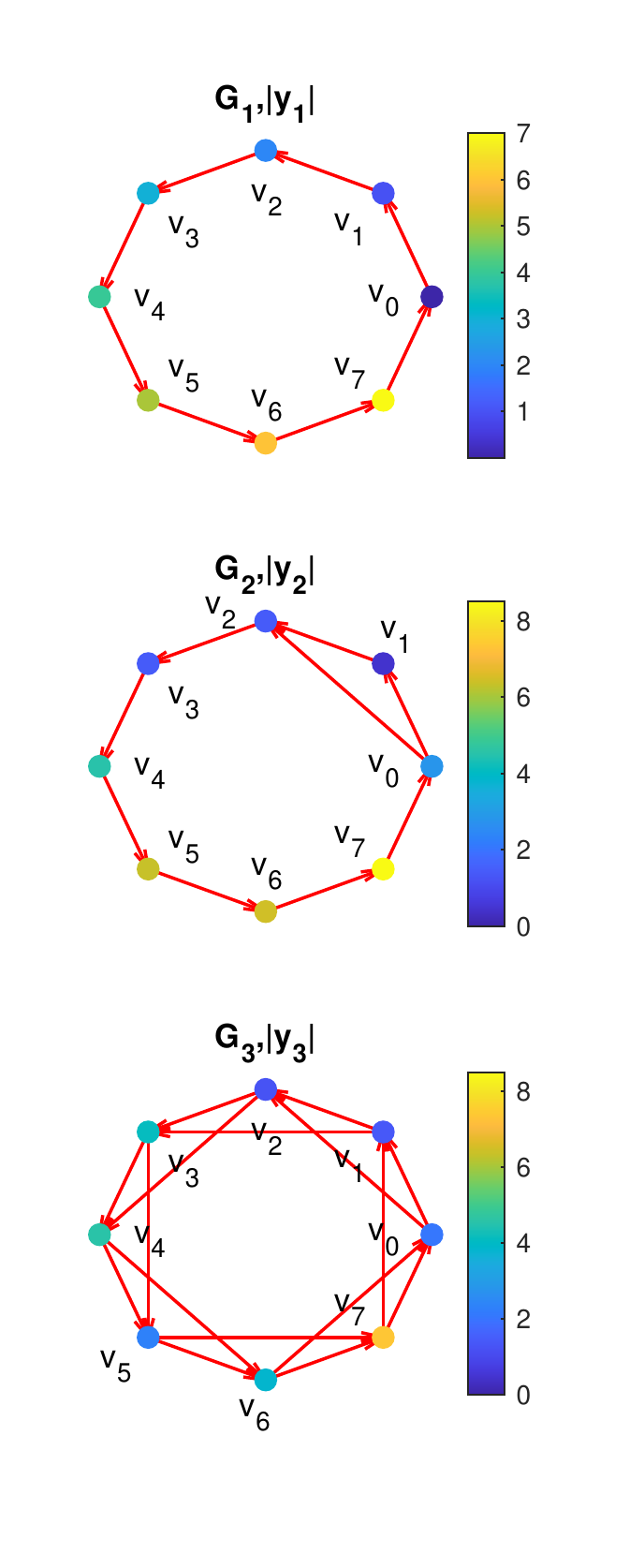} 
  \caption{}
  \label{fig:exampleD}
\end{subfigure}
\caption{Three different graph structures and their associated vertex multiplication operators. (a) The input signal $\mathbf{x} = [1,1,\dots,1]$ plotted on three graphs where colormaps represent signal values. (b)-(c) Real and imaginary parts of vertex multiplication matrices. (d) Magnitudes of the graph signals resulting from applying graph vertex multiplication to the input, i.e. $\mathbf{y_i} = \mathbf{U_{G_i}}\mathbf{x}$.}
\label{fig:example}
\end{figure*}

When viewed as a vector, operators acting on $x$ can be represented by left multiplication with matrices. Of these operators, the one represented by the adjacency matrix itself is of particular importance, since it implicitly contains the connectivity information of $G$. This operator is called a \emph{graph shift} and extends the cyclic shift operator defined on a periodic time series signal in the DSP which has the graph structure shown in Fig.~\ref{fig:exampleA} ($G_1$ at the top part). In this case, the adjacency matrix is identical to the cyclic shift\cite{sandryhaila14freq}:
\begin{equation} \label{eq:tsAdjacency}
    \mathbf{A} = \begin{bmatrix}
                    0& 0&0&...&0&1 \\
                    1& 0&0&...&0&0 \\
                    0& 1&0&...&0&0 \\
                    .& .&.& &.&. \\
                    0& 0&0&...&1&0 \\
                    \end{bmatrix}.
\end{equation}

Let $G$ be a graph with adjacency matrix $\mathbf{A}$ and $x$ be a graph signal defined on $G$. Then $\mathbf{A}$ can be written in the Jordan canonical form as:
\begin{equation}
    \mathbf{A} = \mathbf{V\Lambda V^{-1}}.
\end{equation}
Then, the \emph{Graph Fourier Transform (GFT)} of $x$, denoted by $\Tilde{x}$, is defined as \cite{sandryhaila13discretegsp}:
\begin{equation}
    \Tilde{x} = \mathbf{V^{-1}} x.
\end{equation}
GFT of a signal $x$ is unique up to the ordering of the Jordan blocks in the Jordan canonical form. If the adjacency matrix $\mathbf{A}$ of $G$ is diagonalizable, then the Jordan Canonical form is identical to diagonalizing $\mathbf{A}$. In this case GFT becomes a change of basis into the basis of eigenvectors of $\mathbf{A}$, and this process is unique up to the ordering of the Fourier basis vectors. It is also easy to see that if we are to take the GFT of a time series graph, which has the adjacency matrix of the form given in Eq.~\ref{eq:tsAdjacency}, then the GFT basis becomes the eigenvectors of the cyclic shift matrix, and the GFT reduces to the DFT. 

We now consider the definition of differentiation operation on GSP domain,~\cite{dees19unitaryshift}. Let $x^c: \R \ra \C$ be any smooth periodic function with period $T$, and $\mathbf{t} = (t_0,t_1, ..., t_{N-1})$ be an ordered set of numbers selected from the interval $[0,T)$. Then an irregular sampling of $x^c$ with respect to sampling points $\mathbf{t}$ is the finite discrete signal $x\in\R^N$ where:
\begin{equation}
    x_n = x^c(t_n) \quad \forall n \in \{0,1,\dots,N-1\}.
\end{equation}

Let $S_0$ be the circular forward shift operator defined in the space of finite discrete signals of length $N$. Then the matrix representation of $S_0$ is of the same cyclic shift form given in Eq.~\ref{eq:tsAdjacency}. Then, if we let the addition and subtraction on the vector indices to be defined in modulo $N$ (i.e. if we write $x_{-1} = x_{N-1}$ and $t_{-1} = t_{N-1} - T$), one can use the Taylor series expansion to write:
\begin{align}
    (S_0 x)_n = x_{n-1} &= x^c(t_{n-1}) =
     \sum\limits_{k=0}^{\infty} \bigg(\frac{(t_{n-1} - t_{n})^k}{k!}\frac{d^k}{dt^k}\bigg|_{t_n}x^c\bigg) \nonumber \\  
    &= \bigg(\bigg(\sum\limits_{k=0}^{\infty} \frac{(t_{n-1} - t_{n})^k}{k!}\frac{d^k}{dt^k}\bigg) x^c\bigg)(t_{n}) \nonumber\\ 
    &= \exp{\Big((t_{n-1} - t_{n})\frac{d}{dt}\Big)} x^c(t_{n})  .
\end{align}
Thus, the discrete differential operator $\bm{\nabla}$ should satisfy: 
\begin{equation}
    \mathbf{S_0} = \exp(-\bm{\Delta}_t\bm{\nabla}),
\end{equation}
where $\bm{\Delta}_t = {\rm diag}(t_0 - t_{-1}, t_1 - t_0,...,t_{N-1}-t_{N-2})$.
Then, the matrix manifestation of the differential operator defined on unequal sampling of $x^c$ resulting in the vector $x$ is defined as:
\begin{equation}
\label{diff}
    \bm{\nabla} = -\bm{\Delta}_t^{-1} \log \mathbf{S_0},
\end{equation}
where the complex logarithm can be defined on any branch cut as long as it is consistent throughout the analysis. In this letter, we assume the argument to be in the interval $[0,2\pi)$.

\section{Graph Vertex Multiplication}
Consider a graph $G$ with an adjacency matrix $\mathbf{A}$. Then the eigenvalues of $\mathbf{A}$ can be written as the ordered set $\{r_0e^{j\omega_0},\dots, r_{N-1}e^{j\omega_{N-1}}\}$ where  $r_i$'s are the magnitudes and $\omega_i \in [0,2\pi)$ are given in an increasing order. Then for any graph signal $x$ with a GFT of $\Tilde{x}$, the elements of $\Tilde{x}$ can be ordered with respect to their corresponding eigenvalues. That is, if we write the eigenvalue decomposition of $\mathbf{A}$ as $\mathbf{A = V\Lambda V^{-1}}$ where the diagonal elements of $\mathbf{\Lambda}$ are ordered in an increasing order: $\mathbf{\Lambda} = {\rm diag}(r_0e^{-j\omega_0}, r_1e^{-j\omega_1},...,r_{N-1}e^{-j\omega_{N-1}})$.
Then, $\Tilde{x} = \mathbf{V^{-1}}x$. Notice that we implicitly assume that $\mathbf{A}$ has eigenvalues with distinct arguments (frequencies) when we write the diagonalization of $\mathbf{A}$. In this ordering, each coordinate of $\Tilde{x}$ are in the order of (possibly irregularly) increasing frequency. Since $\Tilde{x}$ has no two coordinates corresponding to the same frequency, we can always find a smooth function $\Tilde{x}^c: \R \ra \C$ such that $\Tilde{x}$ induces an irregular sampling on $\Tilde{x}^c$:
\begin{equation}
    \Tilde{x}_n = \Tilde{x}^c(\omega_n).
\end{equation}
Then the discrete differentiation of $\Tilde{x}$ can be defined as:
\begin{equation}
    (\bm{\nabla}_F\Tilde{x})_n
\end{equation}
where $\bm{\nabla}_F$ is the FT domain discrete differential operator:
\begin{equation}\label{deltaF_old}
    \bm{\nabla}_F = -\bm{\Delta}_\omega^{-1} \log \mathbf{S_0},
\end{equation}
where $\bm{\Delta}_\omega = {\rm diag}(\omega_0 - \omega_{-1}, \omega_1 - \omega_0,..., \omega_{N-1}-\omega_{N-2})$ with $\omega_{-1} = (\omega_{N-1} - 2\pi)$. Before proceeding, to be able to use the precise duality relation given in Eq.~\ref{U_fdf}, we first alter the definition for the discrete derivative by dividing Eq.~\ref{deltaF_old} by $j$:
\begin{equation}
    \Tilde{\bm{\nabla}}_F = j\bm{\Delta}_\omega^{-1} \log \mathbf{S_0},
\end{equation}
which is analogous to the definition of differential operation with constant multiplier $(2\pi j)^{-1}$ as given in Eq.~\ref{eq:operatorD}. (Please note that $2\pi$ term is already encapsulated in frequency $w$.)

Finally, by using the duality in Eq.~\ref{U_fdf}, we can define the precise Fourier dual of $\tilde{\nabla}_F$ as a new operator called \emph{vertex multiplication} denoted by $\mathcal{U_G}$, in abstract operator notation. The matrix manifestation of $\mathcal{U_G}$, using GFT, is then given by:
\begin{equation}
    \mathbf{U}_G = \mathbf{V^{-1}}\tilde{\bm{\nabla}}_F\mathbf{V^{}}=  j\mathbf{V^{-1}}(\bm{\Delta}_\omega^{-1} \log \mathbf{S_0})\mathbf{V}.
\end{equation}

The proposed VM operator $\mathcal{U_G}$ can be interpreted as a collection of vectors $\mathbf{u_i}$ assigned to each vertex on the graph such that $ \mathbf{U}_G = [\mathbf{u}_0,\mathbf{u}_1, \dots,\mathbf{u}_{N-1}]$. Also, it operates on a graph signal $\mathbf{x}=[x_0, \dots ,x_{N-1}]^\top$ as:
\begin{equation} \label{coordinate vector}
   \mathbf{y} = \mathbf{U}_G\mathbf{x} = \sum_{i=0}^{N-1}{x_i\mathbf{u_i}}.
\end{equation}
This definition provides a generalization of the coordinate multiplication operator. As such, the VM operator computes the superposition of the multiplication of the signal value on each vertex with the vector $\mathbf{u_i}$ associated with the same vertex. Hence, the columns of the VM matrix mimic the coordinate values in the coordinate multiplication operator in DSP (Eq.~\ref{eq:operatorU}). Thus, we shall call these columns $\mathbf{u_i}$ the \emph{coordinate vector} of the i'th vertex. Due to the summation in Eq.~\ref{coordinate vector}, the coordinate vector of a vertex has a global effect on the behavior of the VM, and the output graph signal value $\mathbf{y_i}$ depends on values of the input signal at all vertices through the coordinate vectors.

The coordinate multiplication of classical DSP can be interpreted as a special case of VM. The matrix representing the coordinate multiplication, denoted by $\mathbf{U}$, is diagonal. It is composed of one-hot column vectors with the non-zero entries being only at the corresponding indices of each vertex. Then the effect of each coordinate vector in Eq.~\ref{coordinate vector} can be represented locally and thus, the diagonal elements of $\mathbf{U}$ can be assigned to each vertex as proper coordinates. This can be stated more formally in the following lemma:

\textbf{Lemma 1:} For time series graphs, Eq.~\ref{coordinate vector} is equivalent to the coordinate multiplication of the classical DSP.

\textbf{Proof:} Consider the graph representation of a time-series. The adjacency matrix in this case is equivalent to the forward time shift, i.e. $\mathbf{A} = \mathbf{S_0}$. The eigenvalues of $\mathbf{A}$ are then equally spaced on the unit circle in the complex plane and are of the form $e^{j\omega_k}$ where $\omega_k = 2k\pi/N$.
Thus we have
\begin{equation}
\begin{split}
    \bm{\Delta}_\omega &= {\rm diag}(                 \frac{2\pi(0-(-1))}{N}, \frac{2\pi(1-0)}{N},..., \\ & \frac{2\pi((N-1)-(N-2))}{N}) = \frac{2\pi}{N}\mathbf{I},
\end{split}
\end{equation}
where $\mathbf{I}$ is the $N\times N$ identity matrix. Then, the Fourier domain derivative becomes $-\frac{N}{2\pi} \log \mathbf{S_0}$.
This leads to the following vertex multiplication operator:
\begin{equation}
\begin{split}
    \mathbf{U_G} = j\mathbf{V^{-1}}(\frac{N}{2\pi}\log \mathbf{S_0})\mathbf{V} = j\frac{N}{2\pi}\log(\mathbf{V^{-1}} \mathbf{S_0}\mathbf{V)},
\end{split}
\end{equation}
where $\mathbf{V^{-1}}$ reduces to the DFT. Using elementary properties of the DFT, we can obtain
\begin{equation}
\begin{split}
     \mathbf{V^{-1}} \mathbf{S_0}\mathbf{V} = {\rm diag}(e^{-j\omega_0}, e^{-j\omega_1},..., e^{-j\omega_{N-1}})
  = \mathbf{\Lambda}.
\end{split}
\end{equation}
Then, $\mathbf{U_G}$ is the diagonal matrix with entries $\frac{N\omega_k}{2\pi}$. Finally,
\begin{equation}
    \mathbf{U_G} = \mathbf{U} = {\rm diag}(0,1,2,...,N-1),\,
\end{equation}
which is consistent with the discrete coordinate multiplication operator, which can be written as $(\mathbf{U}x)_n = nx_n.$ \hfill $\blacksquare$

\section{Numerical Examples}

\begin{figure}[ht]
\begin{subfigure}{0.24\textwidth}
  \centering
  \includegraphics[width=\linewidth]{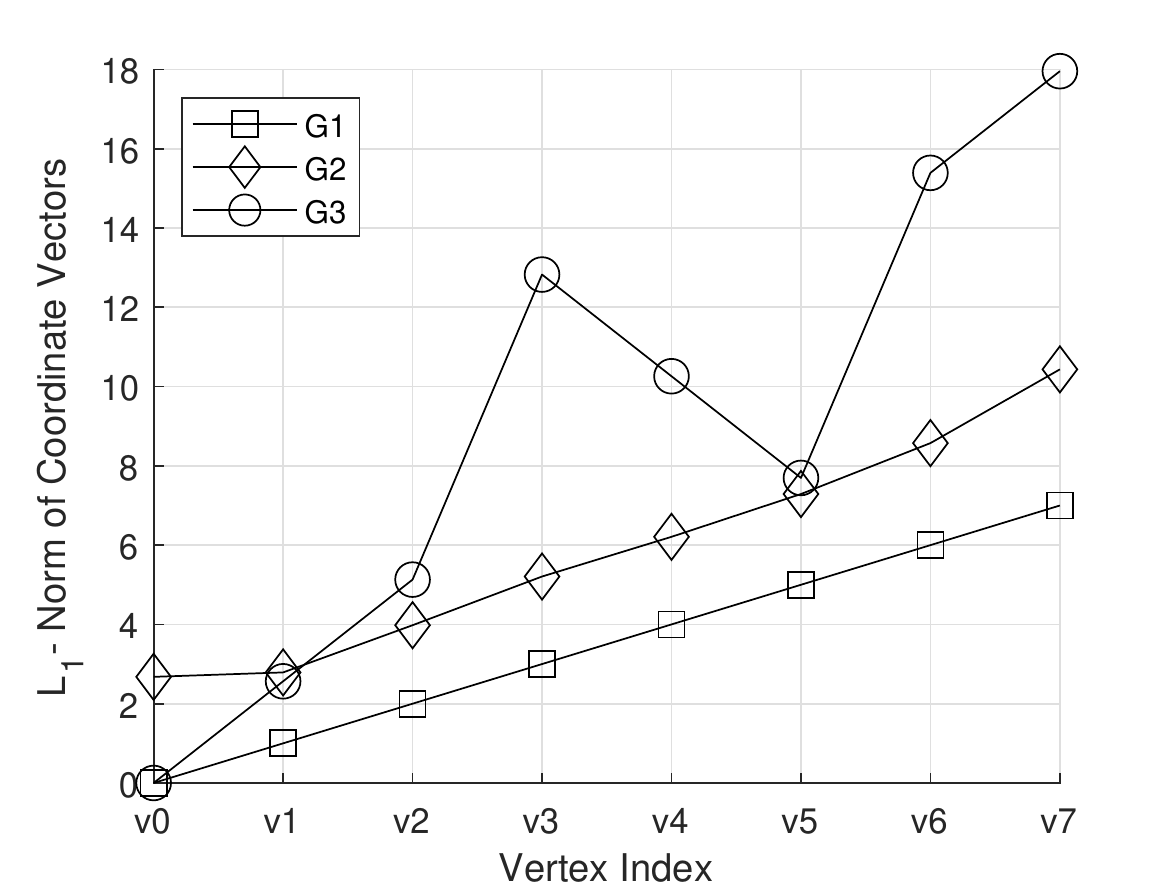} 
  \label{fig:coorA}
\end{subfigure}
\begin{subfigure}{.24\textwidth}
  \centering
  \includegraphics[width=\linewidth]{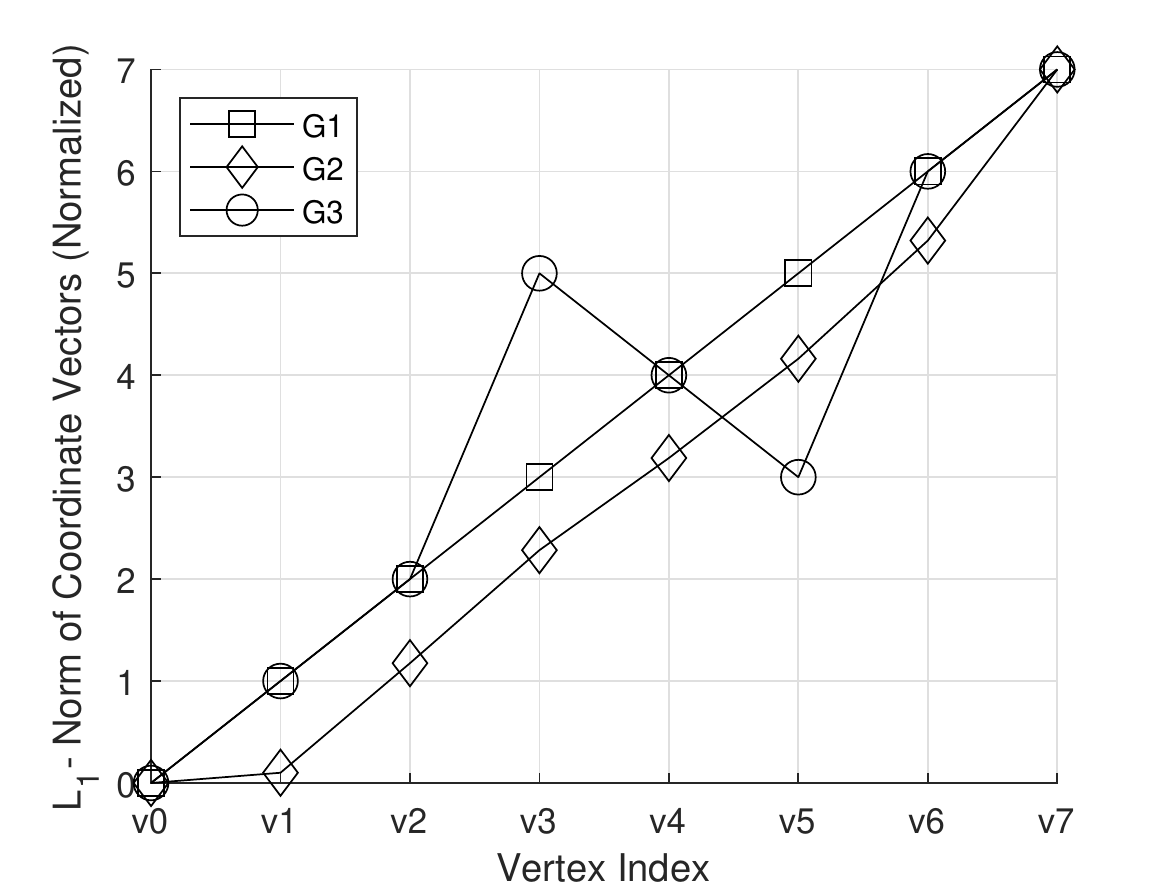} 
  \label{fig:coorB}
\end{subfigure}
\caption{Coordinates are calculated by $L_1$-norm of the coordinate vectors for $G_1$, $G_2$, and $G_3$. (Left: $L_1$-norms. Right: $L_1$-norms normalized to the time-series coordinate interval $[0,7]$.}
\label{fig:coor}
\end{figure}

The effects of the underlying graphs on the corresponding VM operators and their subsequent effects on a graph signal are numerically demonstrated in Fig.~\ref{fig:example}. $G_1$ is the time series graph. We have chosen $G_2$ such that it deviates from time series by only having $v_0$ makes an extra connection to $v_2$; and $G_3$ such that each vertex $v_i$ has connections to the $v_{i+1}$ and $v_{i+2}$. It is immediately clear that any deviation from the time series graph yields to the matrix manifestation of the VM operator to have complex components. Fig.~\ref{fig:example} also presents the vertex-multiplied output graph signals shown by colormaps on the vertices. In fact, experimentation with similar graph structures seems to imply that the largest cycle in the structure has a significant effect on the resulting VM matrix. Further study of this relation can lead to methods for detecting the largest cycles in graph structures. Since VM enables us to define quantitative coordinate vectors, possibilities for manipulations are endless. As an example, in Fig.~\ref{fig:coor}, $L_{1}$-norms of the columns of the VM matrices are also plotted, i.e. given $ \mathbf{U}_G = [\mathbf{u}_0,\mathbf{u}_1, \dots,\mathbf{u}_{N-1}]$, $\mathbf{u}_i$'s are assigned to each vertex and coordinates are calculated by $||\mathbf{u}_i||_1$. Both $L_{1}$-norms and their normalized values to the reference coordinates ($0,1,...,7$) of the time series  by using the scaling $7*\left( ||\mathbf{u}_i||_1 - \min_{i}||\mathbf{u}_i||_1 \right)/\left( \max_{i}||\mathbf{u}_i||_1- \min_{i}||\mathbf{u}_i||_1\right )$ are plotted. Intuitive behavior can be observed in Fig.~\ref{fig:coor}. First, the extra connection between $v_0$ and $v_2$ at the very beginning of the graph almost merges $v_0$ and $v_1$ so that their coordinates become close to each other; then the linearly increasing coordinate structure continues as in the time series. Second, $G_3$ leads to an interesting coordinate pattern due to the regular structure of the graph. Still, a linearly increasing diagonal coordinate structure exists with the abrupt deviation where central vertices $v_3$ and $v_5$ seem to exchange their coordinates.

\section{Conclusions}

We proposed a generalization of the coordinate multiplication operation to the graph domain, called vertex multiplication. By using Fourier duality, the proposed VM is in consistence with the classical signal processing in which differentiation and coordinate multiplication are duals. VM can be interpreted as an operator that assigns a coordinate structure to a graph by assigning each vertex a coordinate vector. These coordinate vectors, which are intrinsically consistent with the FT theory and its dual structure, can be manipulated to further assign single coordinates to the vertices or for other purposes. We showed that the VM reduces to the coordinate multiplication for time series signals. Given an adjacency matrix, such an explicit coordinate association can be helpful in the ongoing generalizations from
classical signal processing to GSP. It may also lead to new theoretical and computational endeavors, and deepen our theoretical understanding of the link between vertex and frequency domains with possible insights and applications to the notion of smoothness, distance metrics and localization in the vertex-domain, and transform designs for signals on graphs.

\ifCLASSOPTIONcaptionsoff
  \newpage
\fi


\bibliographystyle{IEEEtran}

\newpage
\balance

\bibliography{archives_graph,archives_lct}

%

%






\end{document}